\documentclass{article}
\usepackage{graphicx}
\usepackage[T2A]{fontenc}

\newcommand\ov{\over}
\newcommand\ovln{\overline}
\renewcommand\part{\partial}
\newcommand\E[1]{\times10^{#1}}
\newcommand\lesssim{\leq} 
\newcommand\varkappa{\kappa} 

\newcommand\uav{\langle u\rangle}

\newcommand\rs[1]{_\mathrm{#1}}

\newcommand\finj{f\rs{inj}}

\newcommand\fH{f\rs{H}}
\newcommand\fM{f\rs{M}}

\newcommand\me{m\rs{e}}

\newcommand\mpr{m\rs{p}}

\newcommand\nes{n\rs{es}}

\newcommand\nH{n\rs{H}}
\newcommand\nHe{n\rs{He}}

\newcommand\ptail{p\rs{tail}}

\newcommand\pth{p\rs{th}}

\newcommand\rg{r\rs{g}}

\newcommand\vmin{v\rs{min}}

\newcommand\vperp{v_\perp}
\newcommand\vth{v\rs{th}}

\newcommand\yb{y\rs{b}}
\newcommand\ytail{y\rs{tail}}
\newcommand\ymin{y\rs{min}}

\newcommand\Nc{N\rs{c}}

\newcommand\Pcro{P\rs{c}}
\newcommand\Pret{P\rs{r}}

\newcommand\Tes{T\rs{es}}
\newcommand\Teff{T\rs{eff}}

\newcommand\To{T\rs{o}}
\newcommand\Ts{T\rs{s}}

\newcommand\Vs{V\rs{s}}

\newcommand\al{\alpha}
\newcommand\gm{\gamma}

\newcommand\lmb{\lambda}
\newcommand\sg{\sigma}
\newcommand\vsg{\varsigma}
\newcommand\vkp{\varkappa}
\newcommand\Gm{\Gamma}
\newcommand\Dl{\Delta}

\newcommand\chio{\chi\rs{o}}
\newcommand\chis{\chi\rs{s}}

\newcommand\nuesc{\nu\rs{esc}}

\newcommand{\AandA}{A\&A }
\newcommand{\ApJ}{ApJ }
\newcommand{\ApJS}{ApJS }

\begin{document}


\begin{centering}
\noindent{\LARGE\bf Influence of Thermalisation on Electron Injection \\ in Supernova Remnant
Shocks}\\
O. Petruk$^1$ and R. Bandiera$^2$\\
$^1$Institute for Applied Problems in Mechanics and Mathematics, \\
Naukova St. 3-b, Lviv 79000, Ukraine\\
petruk@astro.franko.lviv.ua\\
$^2$Osservatorio Astrofisico di Arcetri, \\ 
Largo E.Fermi 5, Firenze 50125, Italy\\
bandiera@arcetri.astro.it\\
\end{centering}

\vspace{0.5cm}

{\bf Abstract.} {\small
Within a test-particle description of the acceleration process in parallel
nonrelativistic shocks, we present an analytic treatment of the electron
injection.
We estimate the velocity distribution of the injected electrons as the product
of the post-shock thermal distribution of electrons times the probability for
electrons with a given velocity to be accelerated; the injection efficiency is
then evaluated as the integral of this velocity distribution.
We estimate the probability of a particle to be injected as that of going
back to the upstream region at least once.
This is the product of the probability of returning to the shock from
downstream times that of recrossing the shock from downstream to upstream.
The latter probability is
expected to be sensitive to details of the process of electron thermalisation
within the (collisionless) shock, a process that is poorly known.
In order to include this effect, for our treatment we use results of a 
numerical, fully kinetic study, by Bykov \& Uvarov (1999).
According to them, the probability of recrossing depends on physics of
thermalisation through a single free parameter ($\Gamma$), which can be
expressed as a function of the Mach number of the shock, of the level of
electron-ion equilibration, as well as of the spectrum of turbulence.
It becomes apparent, from our analysis, that the injection efficiency is
related to the post-shock electron temperature, and that it results from the
balance between two competing effects: the higher the electron temperature, the
higher the fraction of downstream electrons with enough velocity to return to
the shock and thus to be ready to cross the shock from downstream to upstream;
at the same time, however, the higher the turbulence, which would hinder the
crossing.
}



\section{Introduction}

Strong collisionless shocks are present in various astrophysical objects, and
under a wide range of conditions.
These shocks effectively heat the gas and are also believed to accelerate a
fraction of particles up to very high energies.
Analyses based on multi-frequency observations allow one to
determine properties of both thermal and nonthermal component.
In the case of supernova remnants (hereafter SNRs), optical and UV emission are
typically thermal, radio is nonthermal, while thermal and nonthermal emission
may coexist in X-rays.
The momentum distribution of the accelerated particles is locally well
approximated by a power law; this can be inferred from the power-law
synchrotron spectra, in the case of electrons; while in the case of ions it can
be measured directly in the energy distribution of cosmic rays 
(if one accepts the ``SNR Paradigm'' for the origin of galactic cosmic rays).

Diffusive acceleration (Fermi acceleration) is believed to be the dominant
process that allows particles to gain energies in excess of typically thermal
values.
The standard theory of diffusive acceleration, in test-particle approximation
(see e.g.\ the review of Jones \& Ellison \cite{Jones-Ell-rev-91}), shows that
a power-law distribution develops at high energies.
The spectral index of this high-energy population depends on the shock
compression ratio, while it does not depend at all on the original energy
distribution of the injected particles.
In fact, a simplified way to describe the overall particle evolution, from
nearly thermal to very high velocities, is to treat particle {\it injection}
and {\it acceleration} as two separate problems.
The {\it injection} problem consists in finding out the initial momentum
distribution of that fraction of (originally thermal) particles that can enter
the acceleration process, i.e. to make at least one acceleration cycle. 
The {\it acceleration} problem, instead, consists in following the evolution of
the distribution of these particles along all next acceleration cycles.

In order to keep the number of free parameters low, while modelling more
effectively the emission in all observed spectral ranges, one needs to
introduce a physically self-consistent scenario for the thermal and nonthermal
populations.
For instance, the standard model of particle acceleration constrains the slope
of the electron distribution at high velocities, but does not predict its
normalization: in other terms, the injection efficiency (i.e.\ the fraction of
particles that enter the acceleration process) is poorly known, because this
process is
sensitive to physical details not included in the standard model of Fermi
acceleration.
The level of electron-ion equilibration or, alternatively, the electron
temperature is another key quantity hard
to determine ``a priori'' in collisionless shocks.
While in models of SNR shocks injection and equilibration efficiencies are
taken as independent free parameters, in the reality both depend on the
physical conditions within the shock transition, and therefore they are
not independent.
Goal of this paper is in fact to investigate, in nonrelativistic SNR shocks, a
possible connection between electron injection and thermal equilibration.

A self-consistent treatment of injection and acceleration must include a
microphysical model of particle-wave interactions in the plasma.
A few physical processes have been proposed to account for the electron
injection (see Malkov \& Drury \cite{Malkov-Drury-2001} for a review).
The scattering of electrons is suggested to be due to some ion-generated
instabilities (Bykov \& Uvarov \cite{Byk-Uv-99}; hereafter BU99), whistler 
waves \cite{Levinson-92} and lower-hybrid waves from ions \cite{Leroy-et-al-82}.
These models mainly address the plasma microphysics; while only BU99, to our
knowledge, are able to model the formation of the post-shock electron
distribution.

In this paper, we approach the problem in a simplified way.
We assume the presence of scattering centres, without concentrating on their
nature, but simply assuming that they match the following requirements:
i) the interaction with these scattering centres generates a nearly isotropic,
Maxwellian velocity distribution of particles on timescales not longer than one
collision time; 
ii) the timescales for (wave-mediated) isotropization and energy exchange
between electrons are both smaller than the (wave-mediated) electron-ion
equilibration time;
iii) the scattering centres play at the same time the role of thermalising,
within the shock, the incoming particle population to the post-shock temperature
and that of driving the process of diffusive acceleration.
When we will need to use more specific properties of the wave-particle
interaction (in Sect.~3.2), we will refer to the results of BU99 on the
electron kinetics.

There are in general three ways to calculate the post-shock momentum
distribution of particles: either by solving the kinetic equations, or by
making a hybrid simulation (which is however unable to model the momentum
distribution of electrons because electrons are treated as a fluid), or finally
by extending the individual particle approach of Bell \cite{Bell1978a}.
He has estimated the probability for a particle to
{\em return} to the shock from downstream and has shown that in this way
one obtains a power-law distribution for the accelerated particles with
velocities $v\gg\Vs$ (where $\Vs$ is the shock velocity).

In the present paper we propose to extend the Bell approach to the problem of
injection by introducing the probability to {\em recross} the shock from
downstream to upstream.
This probability is connected to the process of thermalisation of the incoming
flow within the shock.
This fact has been shown by Malkov \cite{malkov-98} in the case of protons.
Namely, the idea that ions are prevented from backstreaming by the
self-generated waves (which also participate in thermalisation of ions)
has allowed Malkov to obtain an analytic solution of the injection problem
for protons.
The main point in his thermal leakage theory is that only those protons that
can ``leak'' upstream are injected into the Fermi process.

We use the same idea in our approach to electrons, even though we are not tight
to any specific kind of interaction.
We consider only the case of parallel shocks, namely when the ambient magnetic
field is parallel to the shock normal.

The plan of the present paper is as follows.
In Sect.~\ref{sect2} we deal with the injection problem by developing an
analytic treatment for calculating the injection efficiency as well as for
determining of the distribution of particles which are able to be accelerated.
Sect.~\ref{sect5} deals with the process of thermalisation and its influence on
injection and, using results from BU99 model, give quantitative estimations on 
the injection efficiency. Sect.~\ref{sect4} concludes.

\section{Injection efficiency and initial distribution}\label{sect2}

\subsection{Efficiency of electron injection}

Let us assume that all electrons are injected into the acceleration process
from the downstream thermal population, i.e.\ we do not invoke seed particles
with velocities already much higher than the thermal velocity.
Their distribution is then well approximated by $\nes\fM$, where
\begin{equation}
\fM(y)={4\ov\sqrt\pi}y^{2}\exp(-y^2)
\label{iso-Maxwel}
\end{equation}
is a normalized Maxwellian, isotropic in the fluid comoving frame.
We have introduced the reduced momentum $y=p/\pth$, which is also equal to the
reduced velocity $v/\vth$, as long as non-relativistic particles are
considered, as it is the case at injection.
Thermal momentum and velocity are defined by $\pth=\me\vth=\sqrt{2\me k\Tes}$,
where $\Tes$ is the post-shock electron temperature.
We consider a fully ionized H+He gas (with $\nHe=0.1\nH$, for a mean mass per
particle $\mu=0.609$), and a strong, unmodified shock.
For an adiabatic index $\gm=5/3$, the shock compression ratio is $\sg=4$ but,
for the sake of generality, in the following formulae we shall allow for a
general $\sg$. Therefore, the ratio of the electron thermal velocity, $\vth$,
to $\Vs$ is
\begin{equation}
 {\vth\over\Vs}=\sqrt{{2(\sg-1)\over\sg^2}{\chis\over\chio}},
 \label{vth-Vs}
\end{equation}
where $\chio=\me/(\mu\mpr)\simeq8.94\E{-4}$.
The factor $\chis=\Tes/\Ts$, where $\Ts$ is the mean shock temperature,
accounts for the thermal equilibration level between electrons and ions
immediately after the shock, and ranges from $\chio$ (no equilibration) to 1
(full equilibration).

Introducing the simple-minded assumption that only particles in the
high-velocity tail of the Maxwellian distribution can be accelerated, it is
easy to link the minimum momentum of this tail, $\ptail=\ytail\pth$, to the
injection efficiency $\vsg$ (i.e.\ the fraction of accelerated particles).
One has just to solve the equation $\int_{\ytail}^{\infty}\fM(y)=\vsg$,
which gives for instance $\ytail=2.85$ for $\vsg=10^{-3}$ and $\ytail=3.91$
for $\vsg=10^{-6}$.
It is worth noticing that, for reasonable values of $\vsg$, this integral is
dominated by particles with $y\sim\ytail$, with $\ytail$ of order of unity:
it is apparent from this example that injection involves mostly particles with
velocities of the order of the thermal one, and not only those with $v\gg\Vs$.

In the above estimation, we have assumed that all particles with $y>\ytail$,
and only them, are accelerated.
In order to find out the injection efficiency in a more general case,
we introduce the probability ${\cal P}(y)$ for a particle with velocity
$v=y\vth$ to be accelerated, i.e.\ to recross the shock from downstream to
upstream at least once.
This probability yields the fraction of particles, with a given velocity, which
can be accelerated; while the Maxwellian distribution in turn gives the number
density of particles with that velocity.
Thus, for an isotropic velocity distribution, the fraction of accelerated
particles (injection efficiency) is given by the integral
\begin{equation}
 \vsg=\int\limits_{0}^{\infty}{\cal P}(y)\fM(y)\,dy.
 \label{varsigm-def}
\end{equation}
In other terms, the distribution of particles injected into the acceleration
process is
\begin{equation}
 \finj(y)={\cal P}(y)\fM(y).
 \label{fini}
\end{equation}
The probability ${\cal P}(y)$ in turn can be estimated as the product of the probability,
$\Pret$, that a particle returns to the shock from downstream, times the
probability, $\Pcro$, that this particle crosses the shock moving upstream.
The next two subsections will be devoted to estimate these two probabilities.

We wish to point out that a common misconception lies underneath the
Fermi acceleration approach, namely that the electrons must enter this
process having already a velocity much higher than $\Vs$.
This is usually obtained, by requiring either
i) that the electron temperature is close to equipartition, or 
ii) that only electrons in the high-energy tail of the Maxwellian distribution
enter into the acceleration process, or finally
iii) that some unknown pre-acceleration mechanism takes place to accelerate
electrons to the required velocity regime.
The condition $v\gg\Vs$ is in fact very useful to simplify the mathematical
treatment of the process, but in our belief is not strictly required by
physical arguments.
In the present paper we will show instead
i) that electrons may be injected efficiently also when their temperatures is
far from equipartition,
ii) that, in order to have reasonably high injection efficiencies, $\chis$ has to be 
considerably less than unity; in other words, the
velocities of the majority of the injected electrons must not be not too far from 
the thermal
velocity and the minimum injection momentum can even be much
smaller than thermal one, and finally
iii) that there is no physical need for an independent pre-acceleration
process, if the treatment of the acceleration is modified in order to account
also for relatively low particle velocities (this can be done by introducing the probability
of crossing the shock, Sect.~\ref{Pcrprob}). 

\subsection{Probability of returning to the shock}

In the case of isotropic velocity distribution in the downstream flow,
the probability for particles with velocity $v$ to return to the shock
from downstream is given by the ratio of the upstream and downstream fluxes
\cite{Jones-Ell-rev-91}:
\begin{equation}
 \Pret(v)={\left|\int\limits_{-v}^{-u_2}(u_2+v_x)\,dv_x\right|\ov
 \int\limits_{-u_2}^{v}(u_2+v_x)\,dv_x}
 =\mathrm{H}(v-u_2)\left({1-u_2/v\ov1+u_2/v}\right)^{2},
 \label{P1-a}
\end{equation}
where: $u_2=\Vs/\sg$ is the velocity of the downstream flow, in the shock
reference frame; $v$ is the velocity, in the downstream flow reference
frame, of the particle that has just reached the shock; $\mathrm{H}(v-u_2)$
is the Heaviside step function (meaning that $u_2$ is the minimum value of $v$
that allows a particle to return to the shock). 
In general, in the paper we label quantities refering to upstream with ``1''
and quantities refering to downstream with ``2''. 
Coordinates are defined in such a way that, in the reference frame of the
shock, the flow moves along the $x$-axis in the positive direction.

It is useful to re-write Eq.~(\ref{vth-Vs}) to fix a lower boundary to the
reduced momenta of the injected particles
\begin{equation}
 \ymin={u_2\ov\vth}=\left({\chio\ov2(\sg-1)\chis}\right)^{1/2}.
 \label{ymin}
\end{equation}
The quantity $\ymin$ is always less than
$(2(\sg-1))^{-1/2}$ (i.e. $p\rs{min}<0.4p\rs{th}$ for $\sg=4$) 
and can be much smaller than that if $\chis\sim1$
(which means $\vth\gg\Vs$).
This means that, the higher the level 
of electron-ion equilibration, the higher the electron thermal velocity
compared to $\Vs$, and thus the higher
the fraction of electrons able to {\em return} to the shock from downstream.

\subsection{Probability of crossing the shock}\label{Pcrprob}

The standard theory of diffusive acceleration \cite{Bell1978a} 
implicitely assumes $\Pcro\simeq1$, which means that the particle mean free
path $\lmb$ is longer than the thickness, $\Dl x$, of the shock transition
region. This condition applies only for particles with high enough velocity 
($v\gg\Vs$).

On the contrary, the evolution of particles with lower velocities is affected
by scatterings within the shock transition. In fact, the mere existence
of a shock implies that the incoming ambient plasma must be thermalised,
within the shock transition region, by some kind of scatterings centres.
Also particles that enter the shock transition region from downstream, as
long as they have velocities similar to thermal particles, must experience
a similar rate of scatterings. Thus, also for them $\lmb<\Dl x$.

In the presence of scatterings, only a fraction of these particles
will succeed crossing the shock and finally reaching the upstream region.
In general, modelling this process is very complex.
Here we will present a simplified treatment, based on some approximations.
The first of them is diffusive approximation, which requires that mean free
paths are smaller than the shock thickness, and that the velocity distribution
is nearly isotropic.

However, this assumption is invalid near the downstream boundary of the shock
layer.
In fact, the original distribution of downstream particles which
return to the shock is highly anisotropic, since all particles entering
the shock have, in the shock reference frame, an $x$-component opposite to
the flow velocity.
The above assumption is anyway valid over most of the volume, provided that
isotropization processes within the shock are very efficient.
Namely, we
require that the length scale for isotropization is of the order of one
mean free path (similarly to what happens for Coulomb collisions between
similar particles).

The estimation of $\Pcro(v)$ is generally very complex. Here we use a crude
approximation (based on the so-called ``modulation'' equation, see e.g.\
 \cite{Jones-Ell-rev-91}) and write
\begin{equation}
 \Pcro=\exp(-\uav\Dl x/\vkp).
 \label{def-a}
\end{equation}
where $\uav\approx(u_1+u_2)/2=u_2(\sg+1)/2$ and $\vkp$ is the diffusion
coefficient.

We want to point out that probability of crossing is closely related to the
thermalization level $\chis$. The thickness $\Dl x$ may be derived from
the condition that the temperature of the incoming fluid increases to the
post-shock value $\Tes=\chis\Ts$ while the fluid moves through the shock
transition from upstream to downstream. In this way, the two problems --
injection and thermalisation -- become closely connected. This can be seen
by rewriting
\begin{equation}
 \Dl x=\uav\Dl t_{12}=
 \uav\int\limits_{0}^{\chis}\left(\frac{d\chi}{dt}\right)^{-1} d\chi
 \label{def-x}
\end{equation}
where $\Dl t_{12}$ is the time it takes to a fluid element to cross
the shock, moving from its upstream boundary to the downstream one. 
In general, it is necessary to introduce a microphysical model of
thermalisation in order to obtain explicitely the functional dependence of
the rate of thermalization $d\chi/dt$.

For the diffusion coefficient we use the standard formula $\vkp=\lmb'v'/3$,
where $v'$ is the velocity of a particle in the local reference frame of the
flow and $\lmb'$ is the particle mean free path with respect to scatterings
within the shock transition.
A further assumption behind this formula is that the scattering centres are
frozen into the fluid. This is, for instance, the case in the BU99 model
for the electron kinetics in a strong shock. In this model, particles are
scattered by the ion-generated Alfv\'enic waves, and the Alfv\'enic
speed is much lower than the shock velocity.
In case of electron diffusion in presence of magnetic field turbulence, it is 
common to
parametrize the mean free path as $\lmb'=\eta\rg$, where $\rg=p'c/eB$
is the gyroradius and $\eta$ accounts for the level of turbulence.
We concentrate here on particles with velocities not much larger than $\Vs$,
and therefore we will use the nonrelativistic formula for $\lmb'$. 
For such parameterization of the mean free path, 
$\lmb'$ may be written as $\lmb'=\tau\rs{D}v'$ where $\tau\rs{D}$ is 
the average deflection time defined as $\tau\rs{D}=\eta m\rs{e}c(eB)^{-1}$.

In order to compute a probability $\Pcro(v)$ for particles having a given
velocity $v$ in the downstream reference frame, we need to average over all
$v'$ velocities corresponding to a a given $v$ downstream.
In the reference frame of the average flow within the shock transition,
velocities $v'$ corresponding to the same $v$ are different in different
directions, namely $\vec{v'}=\vec{v}-(\vec{\uav}-\vec{u_2})$.
The angle-averaged value of $\ovln{v'^2}$ for these particles is given by
\begin{eqnarray}
 \ovln{v'^2}&=&{\int\limits_{-v}^{-u_2}\left((v_x-u_2(\sg-1)/2)^2+\vperp^2
 \right)dv_x\bigg/\int\limits_{-v}^{-u_2}dv_x}\nonumber\\
 &=&v^2+{\sg-1\ov2}vu_2+{\sg^2-1\ov4}u_2^2.
 \label{v2}
\end{eqnarray}

Let us assume that, on the average, electrons in the incoming flow are 
thermalised to the level $\chis$ in $\Nc$ collisionless interactions. 
For the sake of illustration, let us calculate the number of scatterings,
$\Nc$, which yield a given injection efficiency.
The involved time is approximately $\Dl t_{12}=\Nc \tau\rs{D}$ and therefore 
\begin{equation}
 {\uav\Dl x\ov\vkp}={3\uav^2\ov v'^2}{\Dl t_{12}\ov \tau\rs{D}}=
 {3\uav^2\ov v'^2}\Nc,
\end{equation}
so that the probability (\ref{def-a}) becomes
\begin{equation}
 \Pcro(v')=\exp\left(-{3(\sg+1)^2\ov4}\left({u_2\ov v'}\right)^2\Nc\right).
 \label{Pcro}
\end{equation}
Eq.~(\ref{varsigm-def}), together with probabilities (\ref{P1-a}), (\ref{Pcro}) 
and Eqs.(\ref{ymin}), (\ref{v2}), shows that, in order to get an injection efficiency
$\vsg=10^{-3}$,  $\Nc$ must be equal to 9 for $\chis=0.001$, and to 770
for $\chis=0.1$. 

It is interesting to note that our expression for $\Pcro$ behaves like the
``leakage probability'' $\nuesc$ of Malkov \cite{malkov-98}, calculated
for protons.
Namely, the probability for protons to leak across the shock from downstream is
approximately $\nuesc(y)\propto\exp\left(-{\rm const}\ \left(y'\right)^{-2}\right)$
\cite{gies-2000}.

Finally, we want to stress that the introduction of the probability $\Pcro$
does not affect the slope of the accelerated spectrum at relativistic energies.
Following standard test-particle approach to Fermi acceleration 
\cite{Bell1978a}, a power-law momentum distribution of relativistic particles 
is generated, with an
index $\al=-(2-\sg)/(\sg-1)$ that depends only on the shock compression ratio.
This index is obtained by combining the term for the momentum increment per
cycle ($\Dl p/p$) with that for the difference $(1-\Pret)$ per cycle, 
in the high-velocity limit. 
The asymptotic behaviour of both terms is $\propto v^{-1}$; while  
Eq.~(\ref{def-a}) is such that $(1-\Pcro)\propto v^{-2}$. 
Therefore, in the high velocity limit $\Pcro$ gives negligible 
contribution to the formation of the particle spectrum 
in comparison to $\Pret$, and does not affect the formula for $\al$.

\section{Thermalisation of electrons and injection}\label{sect5}

In the shock-front reference frame, if upstream electrons and ions enter the
leading edge of the shock transition with the same velocity then 
the 
electron energy is lower than that of protons by a factor $\me/\mpr$. Therefore, if
the velocities of electrons and ions are randomized independently within the
shock front, we obtain $\Tes=(\me/\mu\mpr)\Ts\ll\Ts$ (i.e.\ $\chis=\chio\ll1$
in our notation), while the ionic temperatures are about $(n\rs{e}+n\rs{i})\Ts/n\rs{i}$
(where ``i'' denotes ions), namely much closer to $\Ts$.

The temperatures of electrons and ions may get closer, if there is a process
within the shock which allows energy exchanges between the two species.
In collisional shocks, the equilibration process is Coulomb scattering
between ions and electrons, while, in the collisionless case (like it generally 
occurs in SNRs), turbulence plays the dominant role.

\subsection{Results from observations}\label{observ}

Since longtime, it has been suggested that plasma instabilities could lead to
prominent heating of electrons within the shock (e.g.\cite{McKee74}).
Some observations and theoretical results put forward the possibility that
collisionless processes
within the shock of SNRs could heat electrons up to the level $\chis\simeq0.4$
(\cite{Bork-Sar-Blond-94} and references therein).
Results on SNR DEM L71 in LMC \cite{Rakows-03} 
and on RCW86 \cite{Ghavam-2001} also suggest $\chis\sim0.3$.
Analysis of Chandra data on Tycho SNR indicates that $\chis\lesssim0.1$ 
\cite{Hwang-et-al-02,Ghavam-2001}.
Other recent observations (SN1006, Tycho, 1E~0102.2--7219) favour a
considerably lower thermalisation level, namely $\chis\lesssim0.03\div0.07$ 
\cite{Korreck-2004,Vink-03,Gvaram-02,Lam-00,Hughes-00,Laming-et-al-96}).

It is important to know how does the level $\chis$ depend on the properties
of the shock. Observational estimations of the shocks with Mach number
${\cal M}$ up to $\sim 400$ suggest that stronger shocks (namely with higher
$V\rs{s}$) could equilibrate species less effectively. Namely, Schwartz
et al. (\cite{Schwartz88}) present results of measurements of $\Tes/\Ts$
for interplanetary shocks and planetary bow shocks (${\cal M}\lesssim 25$)
and find strong evidence that this ratio depends on the Mach number as ${\cal
M}^{-1}$. Ghavamian et al. \cite{Ghavam-2001} estimations for a number of
SNRs seem to extend this trend to stronger shocks, with $25\lesssim {\cal
M}\lesssim 200$. Rakowski \cite{Rakows-trend} summarises the observational
methods and estimations of $\chis$ in SNRs shocks and confirms the inverse
dependence in the range $25\lesssim {\cal M}\lesssim 400$.

\subsection{Results from Bykov and Uvarov (1999)}\label{BUmodel}

Interactions of electrons with ion- or self-generated waves could be
responsible for both accelerating and heating of electrons (see 
\cite{Malkov-Drury-2001,Bykov-2004} for a review).

BU99 have considered the interactions of electrons with ion-generated
electro-magnetic fluctuations and have developed a kinetic model that accounts
at the same time for electron injection, acceleration and thermalisation
in quasiparallel shocks. 
Their model is applicable for shocks with local Mach number 
${\cal M}$ less than $\sim\sqrt{m\rs{p}/m\rs{e}}$. 
They have introduced the effective electron temperature
$\Teff$ (measured in units of the upstream temperature $\To$), which may be
related to our $\chis$ by
\begin{equation}
 \chis=\Teff{\To\ov\Ts}=\Teff{\sg^2\ov(\sg+1){\cal M}^2},
 \label{chi-Teff}
\end{equation}
and have shown that it depends on the Mach number.
This dependence can be approximately described by a power law:
$\Teff\propto{\cal M}^a$, with index $0<a\lesssim2$ depending on which model
of wave-particle interaction is considered.
Therefore, the level of thermalisation depends on the velocity of the shock:
$\chis\propto {\cal M}^{a-2}$ for strong shocks, namely the higher the velocity
the smaller the thermalisation level.

\begin{figure}
\centering
\includegraphics{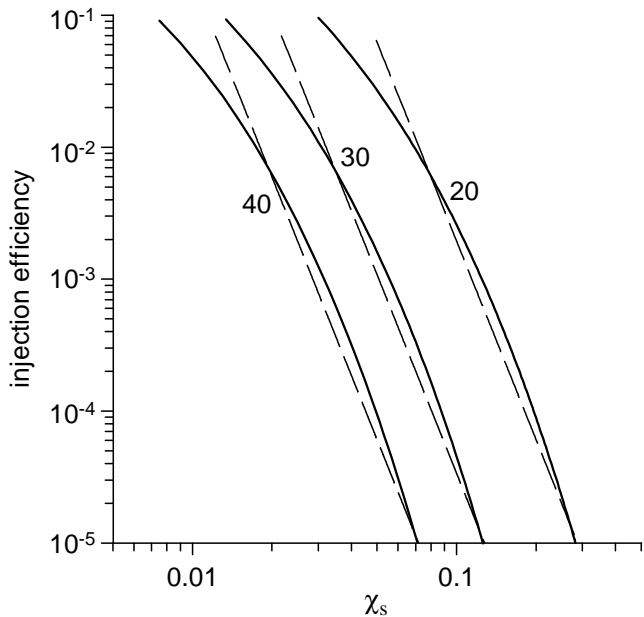}
\caption{Injection efficiency $\vsg$ versus post-shock ratio $\chis$
for a model of BU99 for Bohm-like diffusion and diffusion
boundary conditions. Curves are labelled by their respective Mach number.
The approximation $\vsg=2\E{5}\left({\cal M}^2\chis\right)^{-5}$ is shown by dashed lines.
}
\label{fig-c}
\end{figure}

\begin{figure*}
\centering
\includegraphics{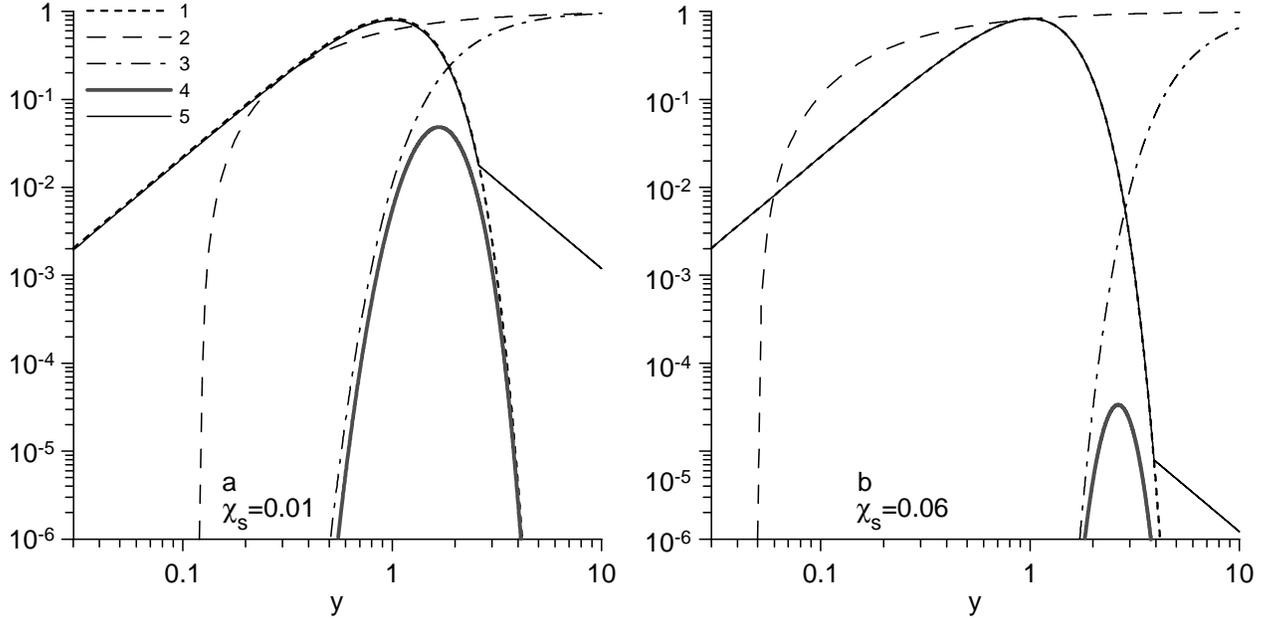}
\caption{Electron distribution functions and probabilities calculated for two different 
values of $\chis$.
1 -- Maxwellian distribution,
2 -- probability to return to the shock $\Pret$,
3 -- probability to cross the shock $\Pcro$,
4 -- initial distribution $\finj$ of electrons injected into acceleration
process,
5 -- final hybrid electron distribution ($\alpha=2$).
Plots are calculated for the model of electron-wave interaction developed
by BU99 (case of Bohm-like diffusion with diffusive boundary conditions), ${\cal
M}=40$.
a)~$\chis=0.01$, in this case $\ymin=0.12$, $\yb=2.6$, $\vsg=0.048$, $\Nc=20$;
b)~$\chis=0.06$, in this case $\ymin=0.05$, $\yb=3.9$, $\vsg=3.1\E{-5}$,
$\Nc=930$.
}
\label{fig-a}
\end{figure*}

If $\lmb'$ is momentum independent (transport of electrons is due to
large-scale magnetic field fluctuations that provide effective heating) then
$a\approx 2$, and the level of equilibration $\chis$ does not depend on the Mach
number. This means that $T\rs{es}\propto \Vs^2$, but with a factor that may be
higher than that inferred from Rankine-Hugoniot equations for the electron population 
\cite{McKee-Hollenbach-1980}.
The opposite case is when electron heating in the shock transition region
is effectively suppressed by a developed small-scale vortex turbulence,
giving $a\approx 0$. In such a situation the postshock electron temperature
is $T\rs{es}\approx T\rs{o}$, independently of the Mach number.
Another interesting model of wave-particle interactions is Bohm-like diffusion,
for which $\lmb'\propto p$ and $a=1$.
In the present paper we consider only the Bohm-like
diffusion case since it seems to
be in agreement with observations (namely $\chis\propto{\cal M}^{-1}$,
see Sect.~\ref{observ}).

BU99 also introduce the dimensionless parameter $\Gm=u_1\Dl x/v'\lmb'$
(calculated for electrons with $v=v\rs{th}$), and in their Fig.~4 they show
its dependence on $\Teff$, for different models of wave-particle interactions
(Fig.~4a for diffusion boundary conditions and Fig.~4b for free escape
boundary conditions). In particular, their curve 4 represents results for
Bohm-like diffusion.

\subsection{Application of BU99 results to our model}\label{BUapplic}

In our paper we have approximated BU99 numerical results
by using $\Gm=\Teff^{1/a}/\xi-1$ where $\xi$ is a constant.
This, together with (\ref{chi-Teff}), gives
\begin{equation}
 \Gm= \left({(\sg+1){\cal M}^2\chis\ov\sg^2}\right)^{1/a}{1\ov\xi}-1.
 \label{Gamma-appr}
\end{equation}
In a Bohm-like case, i.e.\ with $a=1$, $\xi=1.25$ corresponds to the diffusive
boundary conditions and $\xi=0.75$ to free escape boundary conditions.
The parameter $\Gm$ is proportional to the combination $\uav\Dl x/\vkp$ in the
exponent of the transition probability. This allows us to write
\begin{equation}
 \Pcro(y,\chis,{\cal M})=\exp\left(-{3(\sg+1)\ov2\sg}{\Gm(\chis,{\cal M})\ov
y'^{2}}\right)
 \label{Pcro-BU}
\end{equation}
for nonrelativistic electrons and $\lmb'\propto p$. 

By using (\ref{Gamma-appr}) for $\Gm$ in (\ref{Pcro-BU}), the dependence
of the fraction of injected particles $\vsg$ on the level of electron
thermalisation $\chis$ may be obtained (Fig.~\ref{fig-c}). By comparing
(\ref{Pcro-BU}) with (\ref{Pcro}), using (\ref{ymin}) for $y\rs{min}$,
we finally obtain a relation between $\Gm$ and $\Nc$:
\begin{equation}
 \Nc={4(\sg-1)\chis\Gm(\chis)\ov(\sg+1)\sg\chio}.
 \label{nu-BU}
\end{equation}

The calculated dependence of the injection efficiency on the Mach number and
on the thermalisation level is shown in Fig.~\ref{fig-c} for the Bohm-like diffusion. 
The range of
values plotted in this figure corresponds to a range from 2 to 20 for $\Gm$
in Fig.~4 of BU99, with the minimum $\Gm$ corresponding to the maximum $\vsg$.
The curves in Fig.~\ref{fig-c} are essentially the same curve, with different
horizontal offsets.
Namely, the formula for the injection efficiency $\vsg({\cal M},\chis)$ is very
well approximated by a function of ${\cal M}^2\chis$. The reason of this can
be found in Eq.~(\ref{varsigm-def}), together with the explicit definitions of
$\Pret$ and $\Pcro$ (respectively, Eqs.~(\ref{P1-a}) and (\ref{def-a})): for
standard parameter ranges, the most effective term is the argument in the
exponential of $\Pcro$, which is proportional to $\Gm$. In turn,
Eq.~(\ref{Gamma-appr}) shows that the dependence of $\Gm$ from ${\cal M}$ and
$\chis$ is only through the combination ${\cal M}^2\chis$.
In this sense, we may say that $\Gm$ is a function of a single parameter (not
considering, of course, the dependence on the assumed diffusion type and on the
boundary conditions, which can be accounted for by using parameters $a$ and
$\xi$).
For instance, in a Bohm-like case a power-law approximation of the curves
shown in Fig.~\ref{fig-c} is $\vsg\simeq 2\E{5}{\cal M}^{-10}\chis^{-5}$
(this approximation is represented in the figure by dashed lines).
Since in this case $\chis\propto{\cal M}^{-1}$, the overall dependence of
the injection efficiency on the Mach number in a Bohm-like case is as strong
as $\vsg\propto{\cal M}^{-5}$.

In order to allow for different types of diffusion, as well as for different
electron-wave interactions etc., one could consider a more general case,
in which:
i) $\chis\propto {\cal M}^{-m}$; and,
ii) $\vsg\propto{\cal M}^{-2q}\chis^{-q}$ (where we expect $q$ to be always
positive).
We then obtain $\vsg\propto {\cal M}^{-b}$ with $b=q(2-m)$.
In other words, in the case of a decelerating SNR shock $\chis$ always
increases, while $\vsg$ increases if $m<2$, and decreases if $m>2$.

\section{Discussions and Conclusions}\label{sect4}

The electron injection and thermalisation are not independent processes. This
is clearly outlined by Fig.~\ref{fig-a} where the probabilities $\Pret$ and
$\Pcro$ as a function of reduced momentum are shown for two values of $\chis$
together with the initial distribution (\ref{fini}) of injected particles.
The hybrid electron distribution $\nes\fH(y)dy$, Maxwellian up to $\yb$ and
power-law above \cite{Porq-01}, is also shown on the figure
to see the differences.
The break momentum $\yb$ is given by the assumption that all injected particles
obtain momenta higher than $\yb$ after acceleration
i.e. is defined by\ $\vsg=\int_{\yb}^\infty\fH(y)dy$.

\begin{figure}
\centering
\includegraphics{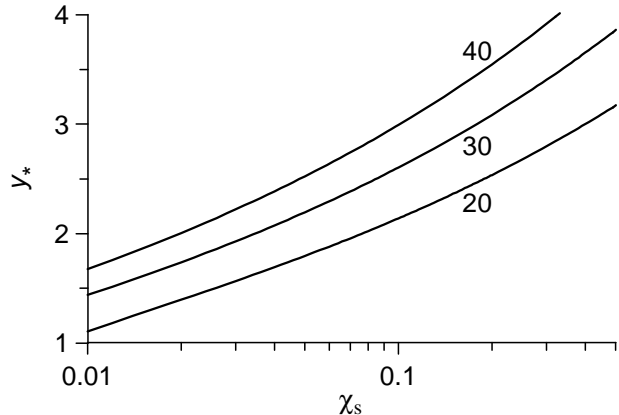}
\caption{Velocity corresponding to the maximum of the distribution function 
$f\rs{inj}$ versus $\chis$ for a few choices of the shock Mach number
(see curve labels).
Model of electron kinetics is the same as in the previous figure.
}
\label{fig-4}
\end{figure}

It is a common believe that only particles from the energetic tail of
Maxwellian distribution are capable to be accelerated.
On the contrary, the distribution $\finj$ shows that thermal particles with
velocities $v>\vmin$ have the possibility to participate in acceleration
process, although with different probability.
The minimum velocity $\vmin=0.07\left(\chis/{0.03}\right)^{-1/2}\vth$ may be 
considerably less than the thermal velocity (see Eq.~(\ref{ymin})). 
The most probable velocity $v\rs{*}$ at which the maximum of the distribution 
$f\rs{inj}$ occurs 
is $v\rs{*}\approx (2\div3)v\rs{th}$ for a wide range of injection fractions
$\vsg=10^{-3}\div10^{-6}$ (Fig.~\ref{fig-4} and compare with Fig.~\ref{fig-c}). 
In other words, most of the electrons are injected with velocities 
$v\rs{*}\simeq 50\chis^{1/2}\Vs$. 

The injection efficiency $\vsg$ of electrons in a collisionless shock is 
associated to the process of electron heating within the shock, through the
competition of two effects.
On one side, the higher the post-shock electron temperature, the higher the
energy of thermal electrons and the higher the fraction of those which are
ready to cross the shock from downstream to upstream (this is given by
the probability $\Pret$, lines 2 on Fig~\ref{fig-a}a,b).
On the other side, however, the higher the temperature, the higher
the number of scattering centers.
Electrons traversing the shock from downstream to upstream also interact
with these sites and the more such interactions the less electrons are
able to cross the shock and to enter into the Fermi acceleration loop (see
probability $\Pcro$, lines 3 on Fig~\ref{fig-a}a,b).
In this paper we show that, for a given ${\cal M}$, the combined effect of
these two processes is that the quantity $\vsg$ decreases with increasing of $\chis$ 
(Fig.~\ref{fig-c}).

Both injection and thermalisation are sensitive to the Mach number. 
It is shown (see Sect.~\ref{BUmodel}) that, for a standard range of
parameters, 
$\vsg(\chis,{\cal M})$ is a decreasing function of a single
argument $\vsg=\vsg({\cal M}^2\chis)$.
Theoretical models show that in high-velocity shocks the
energy of the shock is transferred to the thermal electrons less efficiently,
so that $\chis\propto{\cal M}^{-m}$ with $0\leq m\leq 2$ (BU99).  
Observations favour a dependence
$\chis\propto{\cal M}^{-1}$, suggesting a
Bohm-like type of diffusion.
Our calculations show that the level of electron-ion 
equilibration is expected to depend on the injection fraction as well, so that
the approximate relation between these three parameters is
$\chis\propto {\cal M}^{-2}\vsg^{-1/q}$ 
(Sect.~\ref{BUmodel}). The smaller the Mach number, the higher the level
of electron-ion
equilibration for a given injection efficiency. On the other hand, 
for a given thermalisation level, the stronger the shock the less particles
can be injected.

To conclude, we would like to review the assumptions used in the present paper.

Our approach is in test-particle approximation. Actually, it is known that, in
young SNRs, shocks could be strongly modified; and the inclusion of nonlinear
effects could change our results significantly. The usage of a test-particle
approach in the present paper is however consistent with what done by BU99, and
the results of their analysis are valid at least for shocks with Alfv\'en Mach
number less than $\simeq 43$. Thus our results should be applicable at least to
SNRs either in the late adiabatic phase or beyond. Our
opinion is that, together with using nonlinear treatments, it is valuable
investigating what happens in the linear (test-particle) approximation, also
in consideration that a nonlinear theory has anyway to give, as limit case, the
linear results.

Eqs.~(\ref{iso-Maxwel}) and (\ref{P1-a}) assume nearly isotropic distribution
of particles. In general, this is not fully true for the thermal population
right after the shock.
In order to overcome this difficulty, in numerical calculations these formulae
are assumed to apply a few mean free paths downstream, in order to insure that
the distribution is isotropic in the local frame (e.g. \cite{Ell-et-al-1990}).
Within our approach, this implies some restrictions on the underlying physics.
Our assumptions about properties of the scattering centers in our model (see
Introduction) require that the timescale for isotropisation is not larger than
the timescale for one interaction. 
This means that, in order to assume 
isotropy of particles velocities, we 
would in principle need to increase the number of interactions $N\rs{c}$
at least by one (see Eq.~(\ref{Pcro})).
Since already $N\rs{c}\gg 1$ for, say, $\chi\rs{s}>10^{-3}$ (see estimations
after Eq.~(\ref{Pcro})), this increment would not change much our results, in
the case of shocks producing an electron population thermalised up to the level
$\chi\rs{s}$ higher than $10^{-3}$.
Since $\chis\propto{\cal M}^{-1}$, our models are limited again to the shocks
with moderate Mach numbers.
 
{\small 
{\bf Acknowledgements}
We are grateful to A.~Bykov for helpful discussion on implementation of their
results as well as S.~Reynolds for valuable discussions.
This work has been supported by MIUR under grant Cofin2001--02--10, by MIUR
under grant Cofin2002 and by UFRF 02.07/00430 grant. 
OP acknowledges grant MIUR Cofin2001--02--10 for his fellowship.}


\end{document}